\documentclass[10pt]{article}

\usepackage{amsmath,amsfonts,amssymb,graphicx,epsfig}
\usepackage{psfrag}
\usepackage{colordvi,color}
\title{A set of exactly solvable Ising models with half-odd-integer spin}
\author{Onofre Rojas\thanks{Corresponding author. email: onofre@pq.cnpq.br,  Tel.: +5535 38291954, Fax.: +5535 38291371} and S. M. de Souza \\ Departamento de Ci\^encias Exatas, Universidade Federal de Lavras,\\ CP 3037, 37200-000, Lavras - MG, Brazil.}
\begin{document}

 \maketitle
\begin{abstract}

We present a set of
exactly solvable  Ising models, with half-odd-integer  spin-S on a square-type lattice
including a quartic interaction term in the Hamiltonian. The particular properties of the mixed lattice, 
associated with mixed half-odd-integer spin-(S,1/2) and only nearest-neighbor interaction,
allow us to map this system either onto a purely spin-1/2 lattice
or onto a purely spin-S lattice. By imposing the condition that the mixed half-odd-integer spin-(S,1/2) lattice must have an exact solution, we found a set of exact 
solutions that satisfy the {\it free fermion} condition of the eight vertex model. The number of solutions for a general half-odd-integer spin-S is given by $S+1/2$. Therefore we conclude that
this transformation is equivalent to a simple spin transformation
which is independent of the coordination number. 
\end{abstract}
\sloppy

Two-dimensional lattice models are one of the most interesting subjects of statistical mechanics, 
both experimentally\cite{beg,Bernasconi} and theoretically. Several
approximation methods are used to investigate these  models on the lattice, such as mean-field theory\cite{beg,Capel-Mukamel}, the Bethe
approximation\cite{Chakraborty}, the correlated effective field
theory\cite{Kaneyoshi}, the renormalization group\cite{Krinsky-Berker},
series expansion methods\cite{Soul}, Monte Carlo methods\cite{Jain}
and  cluster variation methods. Following Onsager's\cite{onsager}
solution for the square two dimensional Ising lattice, other solutions
for regular two-dimensional lattices have been considered, such as triangular\cite{newell,husimi},
honeycomb\cite{Horiguchi-Wu,husimi1}, kagom\'e\cite{syozi} lattices and others were
explored in several works and their importance in statistical physics
has waked up the search for a set of completely solvable models.
Some other exact results have been
obtained with restricted parameters, as investigated by Mi and
Yang\cite{MiYang} using a non-one-to-one transformation\cite{Kolesik}.

Some half-odd-integer spin-S Ising models were already discussed in the literature\cite{tang}. 
Using the method proposed by Wu\cite{wu},
Izmailian \cite{izmailian} obtained an exact solution for a spin-3/2 square lattice with only nearest-neighbor and two-body
spin interactions. Izmailian and Ananikian\cite{izm-anani} have also obtained an exact solution for a honeycomb lattice with
spin-3/2. A particular case of
solutions of these models were obtained by the
method proposed by Joseph\cite{joseph}, where any spin-S could be
decomposed in terms of spin-1/2. Another interesting method 
for mapping the
spin-S lattice 
into a spin-1/2 lattice has been proposed by
Horiguchi\cite{horiguchi}.

Mapping between models is an important tool for the study of exactly solvable models.
The aim of this letter is to present the transformation of higher half-odd-integer spin-S systems into a simple spin-1/2 system. 

\begin{figure}[!ht]
\begin{center}
\psfrag{s0}{$S_0$}
\psfrag{s1}{$\sigma_1$}
\psfrag{K}{$K$}
\psfrag{D}{$D$}
\psfrag{La}{${\cal L}_a$}
\psfrag{Lb}{${\cal L}_b$}
\psfrag{Lc}{${\cal L}_c$}
\psfrag{s}{$\sigma$}
\psfrag{m}{$S$}
 \includegraphics[width=16cm,height=7.5cm,angle=0]{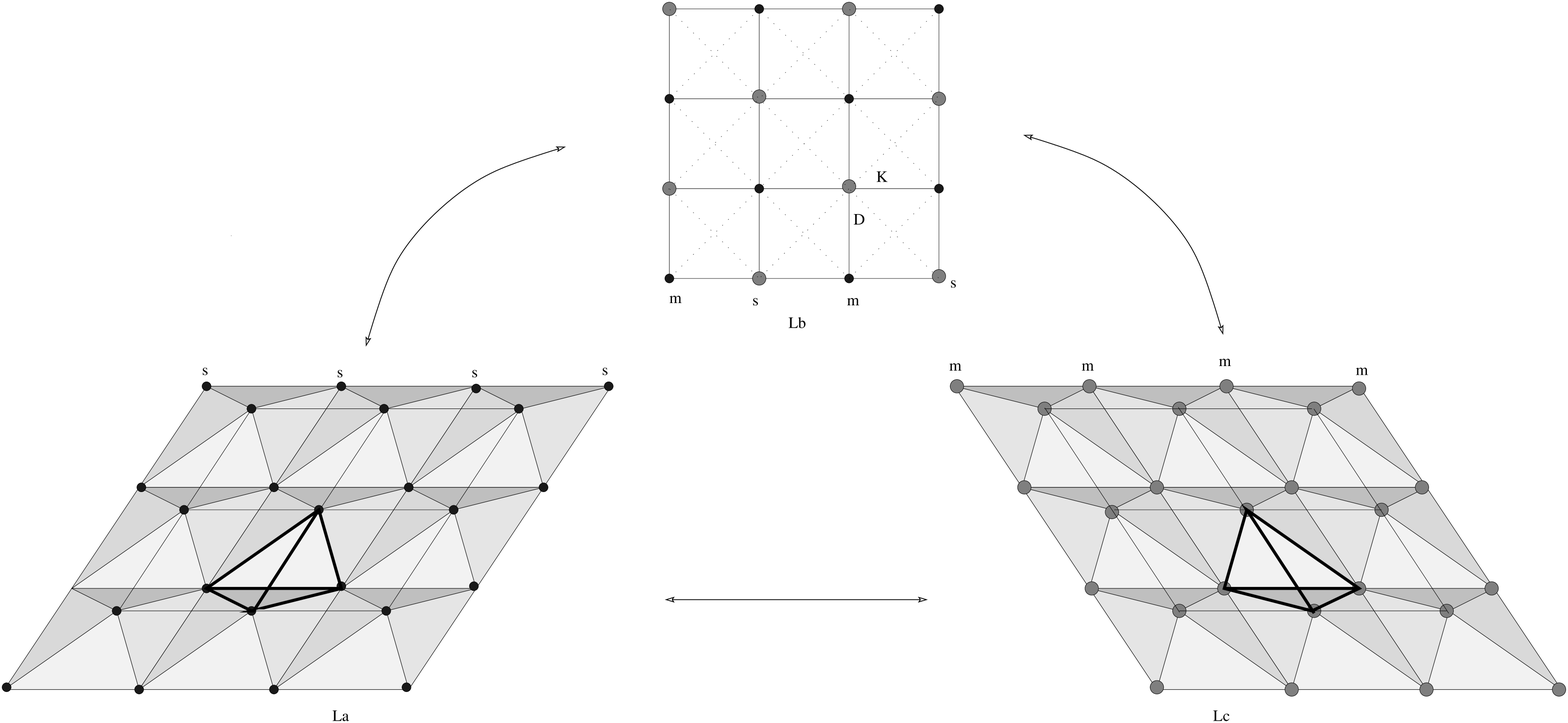}
\end{center}\caption[fig_1]{Schematic representation of mixed
spin-(S,1/2) on a square lattice (${\mathcal L}_b$), a square-type
spin-1/2 Ising model (${\mathcal L}_a$) and a square-type spin-S Ising
model (${\mathcal L}_c$)}
\label{fig_1}
\end{figure}

To demonstrate this transformation we use a two dimensional mixed spin-(S,1/2) on a square Ising lattice with a quartic interaction ${\cal L}_b$ as displayed in fig.\ref{fig_1}. First this mixed model can be mapped onto an exactly solvable two dimensional spin-1/2 Ising model with a quartic interaction\cite{baxter} ${\mathcal L}_a$, such as presented in the literature\cite{izmailian}.  Second it is also possible to transform a mixed spin lattice ${\mathcal L}_b$  onto an effective
spin-S lattice when we consider the spin-1/2 as a decorated Ising model of the
lattice\cite{strecka} ${\cal L}_b$. Then the classical Hamiltonian for a mixed spin-(3/2,1/2) lattice is given by
\begin{align}\label{ising-mixed1}
{\mathcal H}_{1/2,3/2}=\sum_{<i,j>}(K_{r}^{(1)}S_{i}\sigma_{j}+K^{(3)}_{r}S_{i}^3\sigma_{j})+\sum_{i}DS_{i}^2,
\end{align}
with $<i,j>$ meaning summation over nearest interacting neighbors  on the square lattice, and the last summation is performed over all spin-3/2 sites.
The coefficient $K_r^{(1)}$ is the nearest-neighbor interaction parameter of the bilinear term; $K_r^{(3)}$ corresponds to the parameter 
of the non-bilinear interaction, where $r$ runs from 1 up to the coordination
number (in our case up to 4); $D$ is the single ion-anisotropy parameter acting on
spin-3/2; $S_i$ represents the spin-3/2 particle; whereas $\sigma_i$ corresponds the spin-1/2 particle, with two
possible values $\pm 1$ (we use these values conveniently instead of $\pm 1/2$ for all spin $\sigma_i$).

First we would like to obtain an 
exactly solvable spin-(3/2,1/2) lattice $\mathcal{L}_b$ (see fig.1); to this end, we write the first term of the Hamiltonian \eqref{ising-mixed1} as $(K_r^{(1)}S_i+K_r^{(2)}S_i^3)\sigma_j$, then for spin-3/2 we have four possible values $\pm 1/2$ and $\pm 3/2$, thus the term $K_r^{(1)}S_i+K_r^{(2)}S_i^3$ can be written 
\begin{align}\label{As-sol}
A_r^{(1)}=&\pm\tfrac{1}{2}K_r^{(1)}\pm\tfrac{1}{8}K_r^{(3)},\quad S_i=\pm 1/2,\\
A_r^{(2)}=&\pm\tfrac{3}{2}K_r^{(1)}\pm\tfrac{27}{8}K_r^{(3)},\quad S_i=\pm 3/2,\label{As-sol1}
\end{align}
with $r=1,\dots,4$.
 
In order to project the spin-3/2 onto spin $\sigma$ with only two possible values $\pm 1$, we impose the condition $|A_r^{(1)}|=|A_r^{(2)}|=A_r$. Therefore we are able to find the parameters $K_r^{(1)}$ and $K_r^{(3)}$ as a function of $A_r$, by
solving the system \eqref{As-sol} and \eqref{As-sol1}, we obtain following

\begin{eqnarray}\label{K-A}
\left\{
\begin{array}{ll}
K_r^{(3)}=-\frac{4}{3}A_r;& K_r^{(1)}=\frac{7}{3}A_r\\
\text{or}\\
K_r^{(3)}=-\frac{2}{3}A_r;& K_r^{(1)}=-\frac{13}{6}A_r\\
\end{array}
\right., \mbox{ in which } \quad r=1,\dots,4.
\end{eqnarray}

The associated Boltzmann 
weight for a mixed spin-(3/2,1/2) 
lattice has a similar structure to that of the model 
discussed by Wu and Lin\cite{wu-lin}, the associated Boltzmann weights are given by $W(\{\sigma_r\})=\sum_{S}\exp({\mathcal H}_{1/2,3/2})$. For simplicity we consider our calculation in units of $-\beta$. Using the solution given by \eqref{K-A} the associated Boltzmann weights are simplified, which read as
\begin{align}
w_1=&W(+,+,+,+)=\alpha\cosh(A_1+A_2+A_3+A_4),\notag\\
w_2=&W(+,-,+,-)=\alpha\cosh(A_1-A_2+A_3-A_4),\notag\\
w_3=&W(+,-,-,+)=\alpha\cosh(A_1-A_2-A_3+A_4),\notag\\
w_4=&W(+,+,-,-)=\alpha\cosh(A_1+A_2-A_3-A_4),\notag\\
w_5=&W(+,-,+,+)=\alpha\cosh(A_1-A_2+A_3+A_4),\label{boltz-w}\\
w_6=&W(+,+,+,-)=\alpha\cosh(A_1+A_2+A_3-A_4),\notag\\
w_7=&W(+,+,-,+)=\alpha\cosh(A_1+A_2-A_3+A_4),\notag\\
w_8=&W(-,+,+,+)=\alpha\cosh(-A_1+A_2+A_3+A_4),\notag
\end{align}
where $\alpha=\,2({\rm e}^{\frac{D}{4}}+{\rm e}^{\frac{9D}{4}})$. 

The lattice ${\mathcal L}_b$ can be transformed into an effective spin-1/2
lattice ${\mathcal L}_a$, as shown in fig.1. Then at each site of ${\mathcal L}_a$
there can be eight 
lines; we are thus led to consider an eight-vertex
model ${\mathcal L}_a$ with weights given by
\eqref{boltz-w} 
and which satisfies the {\it free fermion} condition\cite{fan-wu},
\begin{align}\label{condition-s}
\omega_1\omega_2+\omega_3\omega_4=\omega_5\omega_6+\omega_7\omega_8.
\end{align}

Consider the following Hamiltonian, 
\begin{align}\label{ising-half}
{\mathcal H}_{1/2}=J_0N+\sum_{(i,j)}J_{i,j}\sigma_{i}\sigma_{j}+\sum_{\substack{\text{all tetra-}\\ \text{hedron}}}J_{1,2,3,4}\sigma_{i_1}\sigma_{i_2}\sigma_{i_3}\sigma_{i_4}
\end{align}
for the effective spin-1/2 Ising model ${\mathcal L}_a$, which is
exactly solvable using the eight-vertex model\cite{baxter}. Here,
$(i,j)$ means summation over all pairs of sites over the tetrahedron (the
boldface tetrahedron in Fig.1), and $J_{i,j}$ are their corresponding
interacting parameters, whereas $J_{1,2,3,4}$ is the quartic
interaction 
parameter; $N$ represents the number of spin-3/2 sites on the
lattice. 
 
After 
transforming the Hamiltonian \eqref{ising-mixed1} into \eqref{ising-half}, we relate their parameters
using the Boltzmann weights, thus to obtain
\begin{align}\label{J_0}
J_0=&\ln \big(2\alpha\big)+\frac{1}{8}\ln\big(w_1w_2w_3w_4w_5w_6w_7w_8\big)\\
J_{1,2}=&\frac{1}{8}\ln\Big(\frac{w_1w_4w_5w_8}{w_6w_7w_2w_3}\Big)\\
J_{2,3}=&\frac{1}{8}\ln\Big(\frac{w_1w_8w_3w_6}{w_7w_4w_5w_2}\Big)\\
J_{3,4}=&\frac{1}{8}\ln\Big(\frac{w_1w_6w_7w_4}{w_5w_2w_3w_8}\Big)\\
J_{4,1}=&\frac{1}{8}\ln\Big(\frac{w_1w_7w_5w_3}{w_8w_4w_2w_6}\Big)\\
J_{1,3}=&\frac{1}{8}\ln\Big(\frac{w_1w_7w_2w_8}{w_6w_4w_5w_3}\Big)\\
J_{2,4}=&\frac{1}{8}\ln\Big(\frac{w_1w_6w_5w_2}{w_7w_4w_3w_8}\Big)\\ \label{J_1234}
J_{1,2,3,4}=&\frac{1}{8}\ln\Big(\frac{w_1w_4w_2w_3}{w_8w_7w_5w_6}\Big).
\end{align}

The partition function ${\mathcal Z}$ 
of the decorated (or mixed) model in
the thermodynamic limit is related to the partition function
${\mathcal Z}_{8v}$ of the effective eight-vertex spin-1/2 model
\eqref{ising-half} by the expression ${\mathcal Z}={\rm e}^{-\beta
J_0}{\mathcal Z}_{8v}$. 
An analytical expression for the free energy of the
{\it free fermion} model is well known\cite{fan-wu} and after some
manipulation, was expressed  in the thermodynamic limit by
\begin{align}
f=J_0+\frac{1}{16\pi^2}\int_0^{2\pi}\int_0^{2\pi}\,{\rm d}\theta{\rm d}\phi\ln[2a+2b\cos(\theta)+2c\cos(\phi)+2d\cos(\theta-\phi)+2e\cos(\theta+\phi)],
\end{align}
where
\begin{align}
a=&\tfrac{1}{2}(w_1^2+w_2^2+w_3^2+w_4^2),& b=&w_1w_3-w_2w_4,&\notag\\
c=&w_1w_4-w_2w_3,& d=&w_3w_4-w_7w_8,& e=&w_3w_4-w_5w_6.
\end{align}

The system exhibits an Ising transition at the critical points
\begin{align}
w_1+w_2+w_3+w_4=2\ \text{max}\{w_1,w_2,w_3,w_4\}.
\end{align}
At low temperature the system exhibits ordered states  
such as 
the ferromagnetic state (for $w_1>w_2,w_3,w_4$),
the antiferromagnetic state (for $w_2>w_1,w_3,w_4$)
and the metamagnetic state (for $w_3$ or $w_4>w_1,w_2$).

 Thus our goal is to transform an 
exactly solvable Ising model with spin-1/2 via the eight-vertex
model\cite{baxter} into an equivalent Ising model on the lattice with
spin-3/2. For this purpose we introduce an auxiliary lattice ${\cal
L}_b$, with mixed spin-1/2 and spin-3/2, the schematic transformation of which 
is displayed in fig.\ref{fig_1}. If we consider the spin-1/2 model as a decoration of the mixed spin model and transform it into an equivalent spin-3/2 Ising model on lattice, then
we can conclude that 
there exist a transformation 
from the spin-1/2 Ising
model onto the spin-3/2 Ising model, with a non-bilinear
interaction and four-body interactions terms over the tetrahedron or
quartic interaction.

Using the first  of solution of Eq.\eqref{K-A}, we express  the
Hamiltonian of the transformed spin-3/2 lattice $\mathcal L_c$ in terms of
four constrained parameters and one arbitrary parameter $D$, thus obtaining
\begin{align}\label{ising-3/2}
{\mathcal H}_{3/2}=&J_0N+\sum_{i}DS_i^2+\sum_{(i,j)}J_{i,j}\big(\frac{49}{9}S_{i}S_{j}-\frac{28}{9}(S_{i}S_{j}^3+S_{i}^3S_{j})+\frac{16}{9}S_{i}^3S_{j}^3\big)+\sum_{\substack{\text{all tetra-}\\ \text{hedron}}}J_{1,2,3,4}\Big(\frac{2401}{81}S_{i_1}S_{i_2}S_{i_3}S_{i_4} \notag\\& -\frac{1372}{81}S_{i_1}^3S_{i_2}S_{i_3}S_{i_4}+ \frac{784}{81}S_{i_1}^3S_{i_2}^3S_{i_3}S_{i_4}-\frac{448}{81}S_{i_1}S_{i_2}^3S_{i_3}^3S_{i_4}^3 +\frac{256}{81}S_{i_1}^3S_{i_2}^3S_{i_3}^3S_{i_4}^3\Big),
\end{align}
where the parameters $J_{i,j}$ were already defined by 
Eqs.\eqref{J_0}-\eqref{J_1234}, and these parameters are constrained. Whereas the parameters $A_1$, $A_2$, $A_3$
and $A_4$ are free, as well as the single ion anisotropy parameter
$D$. Summations are performed as indicated in \eqref{ising-half}.

By factorizing the terms under the summation sign, we obtain
\begin{align}\label{ising-3/2-r}
{\mathcal H}_{3/2}=&J_0N+\sum_{i}DS_i^2+\sum_{(i,j)}\big(J_{i,j}\frac{S_{i}S_{j}}{9}(7-4S_{i}^2)(7-4S_{j}^2)\big)+\notag\\ &+\sum_{\substack{\text{all tetra-}\\ \text{hedron}}}J_{1,2,3,4}\frac{S_{i_1}S_{i_2}S_{3}S_{i_4}}{81}(7-4S_{i_1}^2)(7-4S_{i_2}^2)(7-4S_{i_3}^2)(7-4S_{i_4}^2).
\end{align}

We can reduce the Hamiltonian \eqref{ising-3/2-r} further for both 
solutions in Eq.\eqref{K-A}, using the spin
transformation $\sigma^{(k)}(S)$, where this function only take two
possible values $\pm 1$, for all values of
$S=\{-3/2,-1/2,1/2,3/2\}$. These spin transformations read as
\begin{eqnarray}\label{trnsf-3/2}
\sigma^{(k)}(S) = 
\left\{
\begin{array}{lr}
\frac{S}{3}(7-4S^2);& k=1\\
\frac{S}{6}(13-4S^2);& k=2.\\
\end{array}
\right.
\end{eqnarray}
We remark 
that the transformation given by \eqref{trnsf-3/2}
are the same as those obtained by Izmailian\cite{izmailian} using the method proposed by Wu\cite{wu} (but for another model, the spin-3/2 Ising model on a square
lattice).  As we can see, the previous transformation is independent of the coordination number.
Therefore it may be used to yield 
a particular case of solution, for 
the exactly solvable spin-3/2 Ising model in a honeycomb
lattice as was obtained by Izmailian and Ananikian\cite{izm-anani}.

We are now able to write the Hamiltonian \eqref{ising-3/2-r}
as
\begin{align}\label{Ham-3/2}
{\mathcal H}^{(k)}_{3/2}=&J_0N+\sum_{i}DS_i^2+\sum_{(i,j)}J_{i,j}\sigma_i^{(k)}\sigma_j^{(k)}+\sum_{\substack{\text{all tetra-}\\ \text{hedron}}}J_{1,2,3,4}\sigma_{i_1}^{(k)}\sigma_{i_2}^{(k)}\sigma_{i_3}^{(k)}\sigma_{i_4}^{(k)}.
\end{align}

 The second term of the Hamiltonian \eqref{Ham-3/2} may be written
using a new spin variable transformation, similar to that performed by Izmailian\cite{izmailian}. Note that there the transformation was performed only for spin-3/2 case and for higher spin that method could become a more complex task.

 We also
remark 
that the model discussed by Izmailian\cite{izmailian} can be
completely re-obtained, by 
using
our simple transformation
instead of Wu's\cite{wu} method.

The Hamiltonian \eqref{ising-mixed1} can be
extended to arbitrary half-odd-integer spin-(S,1/2); the general
Hamiltonian has the following form
\begin{align}\label{ising-mixedS}
{\mathcal H}_{1/2,S}=\sum_{<i,j>}\Big(K_{r}^{(1)}S_{i}+K^{(3)}_{r}S_{i}^3+K^{(5)}_{r}S_{i}^5\dots +K^{(S+\frac{1}{2})}_{r}S_{i}^{S+\frac{1}{2}}\Big)\sigma_{j}+\sum_{i}DS_{i}^2,
\end{align}
where $K_r^{(1)}$ is the bilinear interaction parameter, and
$K^{(u)}_{r}$ are  non-bilinear interaction parameters of
$S_i^u\sigma_j$, with $u=2,...,S/2+1$ and $r=1,...,4$, thus the set of
transformations for each half-odd-integer spin-S have $S+1/2$ unknown parameters
to be determined. We  extend 
eq.\eqref{K-A} to the general
case \eqref{ising-mixedS} 
by constructing a Vandermonde-like matrix, explicitly

\begin{align}
\begin{pmatrix}
\frac{1}{2}& (\frac{1}{2})^3& (\frac{1}{2})^5& \dots& (\frac{1}{2})^{2S}\\
\frac{3}{2}& (\frac{3}{2})^3& (\frac{3}{2})^5& \dots& (\frac{3}{2})^{2S}\\
\vdots&\vdots &\vdots  &\ddots  &\vdots \\
S& S^3& S^5&  \dots& S^{2S}
\end{pmatrix} 
\begin{pmatrix}
K_r^{(1)}\\ K_r^{(3)} \\ \dots\\ K_r^{(S+\frac{1}{2})}
\end{pmatrix}=\begin{pmatrix}
A_r^{(1)}\\ 
A_r^{(2)} \\ \vdots\\ A_r^{(S+\frac{1}{2})}
\end{pmatrix} ,
\end{align}
where the parameters $A_r^{(u)}$ must satisfy the following identities
$|A_r^{(1)}|=|A_r^{(2)}|=\dots=|A_r^{(S+\frac{1}{2})}|=A_r$, in order to project the spin-S onto spin-1/2.
By inverting the Vandermonde-like matrix, we are able to
obtain the solution of the algebraic system. In what follows, we show
the solutions of 
these system of linear equations for some higher spin-S values.

For $S=5/2$, there are three solutions:
\begin{eqnarray}\label{trnsf-5/2}
\sigma^{(k)}(S) = 
\left\{
\begin{array}{ll}
{\frac {1067}{480}}\,S-{\frac {11}{12}}\,{S}^{3}+\frac{1}{10}\,{S}^{5};& k=1\\
{\frac {529}{240}}\,S-\frac{5}{6}\,{S}^{3}+\frac{1}{15}\,{S}^{5};& k=2\\
{\frac {1183}{480}}\,S-{\frac {23}{12}}\,{S}^{3}+{\frac {7}{30}}\,
{S}^{5}; & k=3.
\end{array}
\right.
\end{eqnarray}
For $S=7/2$, there are four solutions:
\begin{eqnarray}\label{trnsf-7/2}
\sigma^{(k)}(S) = 
\left\{
\begin{array}{ll}
{\frac {30251}{13440}}\,S-{\frac {301}{288}}\,{S}^{3}+{\frac {61}{
360}}\,{S}^{5}-{\frac {1}{126}}\,{S}^{7}
;& k=1\\
{\frac {60577}{26880}}\,S-{\frac {3047}{2880}}\,{S}^{3}+{\frac {
127}{720}}\,{S}^{5}-{\frac {11}{1260}}\,{S}^{7}
;& k=2\\
{\frac {14887}{6720}}\,S-{\frac {637}{720}}\,{S}^{3}+{\frac {17}{
180}}\,{S}^{5}-{\frac {1}{315}}\,{S}^{7}
; & k=3\\
{\frac {68123}{26880}}\,S-{\frac {1289}{576}}\,{S}^{3}+{\frac {293
}{720}}\,{S}^{5}-{\frac {5}{252}}\,{S}^{7};& k=4.
\end{array}
\right.
\end{eqnarray}
For $S=9/2$, we obtain five solutions:
\begin{eqnarray}\label{trnsf-9/2}
\sigma^{(k)}(S) = 
\left\{
\begin{array}{ll}
{\frac {5851067}{2580480}}\,S-{\frac {46573}{41472}}\,{S}^{3}+{
\frac {7501}{34560}}\,{S}^{5}-{\frac {97}{6048}}\,{S}^{7}+{\frac {
1}{2592}}\,{S}^{9};& k=1\\
{\frac {2924921}{1290240}}\,S-{\frac {813413}{725760}}\,{S}^{3}+{
\frac {3727}{17280}}\,{S}^{5}-{\frac {239}{15120}}\,{S}^{7}+{
\frac {17}{45360}}\,{S}^{9};& k=2\\
{\frac {5868067}{2580480}}\,S-{\frac {334907}{290304}}\,{S}^{3}+{
\frac {8117}{34560}}\,{S}^{5}-{\frac {113}{6048}}\,{S}^{7}+{\frac 
{43}{90720}}\,{S}^{9}; & k=3\\
{\frac {5725183}{2580480}}\,S-{\frac {37337}{41472}}\,{S}^{3}+{
\frac {3593}{34560}}\,{S}^{5}-{\frac {29}{6048}}\,{S}^{7}+{\frac {
1}{12960}}\,{S}^{9};& k=4\\
{\frac {6651283}{2580480}}\,S-{\frac {506017}{207360}}\,{S}^{3}+{
\frac {18341}{34560}}\,{S}^{5}-{\frac {1237}{30240}}\,{S}^{7}+{
\frac {13}{12960}}\,{S}^{9}; & k=5.
\end{array}
\right.
\end{eqnarray}

Using the spin transformations given by eqs.\eqref{trnsf-3/2},
\eqref{trnsf-5/2}, \eqref{trnsf-7/2} and \eqref{trnsf-9/2}, we are
able to recover the transformation obtained by Joseph\cite{joseph}
by considering $k=1$, whereas the remaining transformation are new.


In this report we present 
a transformation which maps a  spin-3/2
lattice, with quartic  and non-linear interactions terms,
onto an effective spin-1/2  Ising model on lattice. 
First this transformation was
carried out using an auxiliary mixed spin-(3/2,1/2) square
lattice with only nearest neighbor interaction term, 
mapped onto an effective 
spin-1/2 or  spin-3/2 lattice model, depending on which spin is considered to be the decoration spin. 
  Finally, a systematic way of transformation for higher half-odd-integer spin-S is considered, inverting a Vandermonde like matrix, to obtain a families of mapping between spin-S models and spin-1/2 models, using the one-to-one transformation. Therefore we conclude that there exist a spin transformation, which can be applied to lattice models with arbitrary coordination number,  including non-exactly solvable half-odd-integer spin-S models. 
We also recovered some results previously obtained in the literature\cite{izmailian,izm-anani}, as a particular case of solution.

 O. Rojas. and S.M. de Souza. thanks CNPq and FAPEMIG for partial financial support.

\newpage

\end{document}